\documentclass[twocolumn,twoside,preprintnumbers,amsmath,amssymb,showkeys]{revtex4}
\usepackage{epsfig}
\usepackage{graphicx}

\usepackage{fancyhdr}

\usepackage{pslatex}

\def\met{$E\!\!\!\!\!/_T$}
\def\etjet{$E_T^{jet}$}

\def\w1jet{$W+1{\rm -jet}$ }
\def\wevnjet{$W\to e\nu +n{\rm -jet}$ }

\def\wgenjet{$W+\geq n{\rm -jet}$ }
\def\wge1jet{$W+\geq 1{\rm -jet}$ }

\def\wev{$W\to e\nu$ }

\begin{document}

\title{Measurement of the W + jet cross section at CDF}

\author{Andrea Messina, on behalf of the CDF collaboration}

\affiliation{Michigan State University, 
         3218 Biomedical Physical Science, East Lansing, MI 48824-2320 USA}


\begin{abstract}

A measurement of \wevnjet cross sections in $p\bar{p}$ collisions at 
$\sqrt{s}=1.96$~${\rm TeV}$ using the Collider Detector at Fermilab in Run II
is presented. The measurement is based on an integrated luminosity of $320$~${\rm pb^{-1}}$, 
and includes events with jet multiplicity from $\ge 1$ to $\ge 4$. In each jet
multiplicity sample the differential and cumulative cross sections with respect to 
the transverse energy of the $n^{th}-$leading jet are measured. 
For $W+\ge2$ jets the 
differential cross section with respect to the 2-leading jets invariant mass $m_{j_{1}j_{2}}$
and angular separation $\Delta R_{j_{1}j_{2}}$ is also reported.
The data are compared to predictions from Monte Carlo simulations.

\end{abstract}
\maketitle

\thispagestyle{fancy}

\setcounter{page}{1}

The study of jets produced in events containing a $W$ bosons provides 
a useful test of Quantum Chromo-Dynamics (QCD) at high momentum transfers.
Recently a lot of work ~\cite{ref:MCLO} has been invested to develop 
sophisticated Monte Carlo programs  capable of handling more particle in the final
state at the leading order (LO), or in some cases, next-to-leading order
(NLO) in perturbative QCD. 
Measurements of W + jet cross sections are an important test of QCD and
may be used to validate these new approaches. A good understanding of W + jet
production is vital to reduce the uncertainty on the background to top pair production
and to increase the sensitivity to higgs and new physics searches at the Tevatron
and the LHC.

This contribution describes a measurement of the $W\to e \nu +\geq n{\rm -jet}$  
production cross section in $p\bar{p}$ collisions at a center of mass energy of 1.96 TeV.
The cross section is presented for four inclusive n-jets samples 
($n=$ 1, 2, 3, 4) as a function of the $n^{th}-$leading jet transverse energy ($E^{jet}_T$). 
For $W+\ge2$ jets the 
differential cross section with respect to the 2-leading jets invariant mass $m_{j_{1}j_{2}}$
and angular separation $\Delta R_{j_{1}j_{2}}$ is also reported.
%
Cross sections have been corrected to particle level jets, and are defined within a limited W
decay phase space, closely matching that which is experimentally accessible. 
This definition, easily reproduced theoretically, minimizes the model dependence that can
enter a correction back to the full W cross-section. These results thus offer the potential
for extensive tuning of W+jet(s) Monte Carlo approaches at the hadron-level.
This analysis is based on $320 \pm 18$~${\rm pb^{-1}}$ of data 
collected by the upgraded Collider Detector at Fermilab (CDF II) during the Tevatron Run II
period.

The CDF II detector~\cite{ref:CDF} is an azimuthally and forward-backward symmetric
apparatus situated around the $p\bar{p}$ interaction region, consisting of a magnetic
spectrometer surrounded by calorimeters and muon chambers. 

\wev candidate events are selected from a high $E_T$ electron trigger ($E^e_T\ge18$ GeV, $|\eta^e|<1.1$) 
by requiring one good quality electron candidate ($E^e_T\ge20$ GeV) and the missing transverse energy 
(\met) to be greater than $30$~${\rm GeV}$. 
To further reduce background contamination, the $W$ transverse mass is required
to satisfy $m_T^W>20$~${\rm GeV/c^2}$. In addition, $Z\to e^{+}e^{-}$  are rejected with a veto 
algorithm designed to identify event topologies consistent with having a 
second high $E_T$ electron.
The \wev candidate events are then classified according to their jet multiplicity into
four inclusive $n-$jet samples ($n=1,~4$). 
Jet are searched for using an iterative seed-based cone algorithm~\cite{ref:jetalg}, 
with a cone radius $R=\sqrt{(\Delta\eta)^2+(\Delta\phi)^2}=0.4$.
Jets are requested to have a corrected transverse energy $E_T^{jet}>15$GeV and a pseudorapidity
$|\eta|<2.0$. 
$E_T^{jet}$ is corrected on average for the
calorimeter response and the average contribution to the jet energy from additional $p\bar{p}$
interaction in the same bunch crossing~\cite{ref:jetnim}. No correction is applied for the 
contribution to the jet energy coming from the underlying event. 

\begin{figure}
\includegraphics[width=.181\textheight]{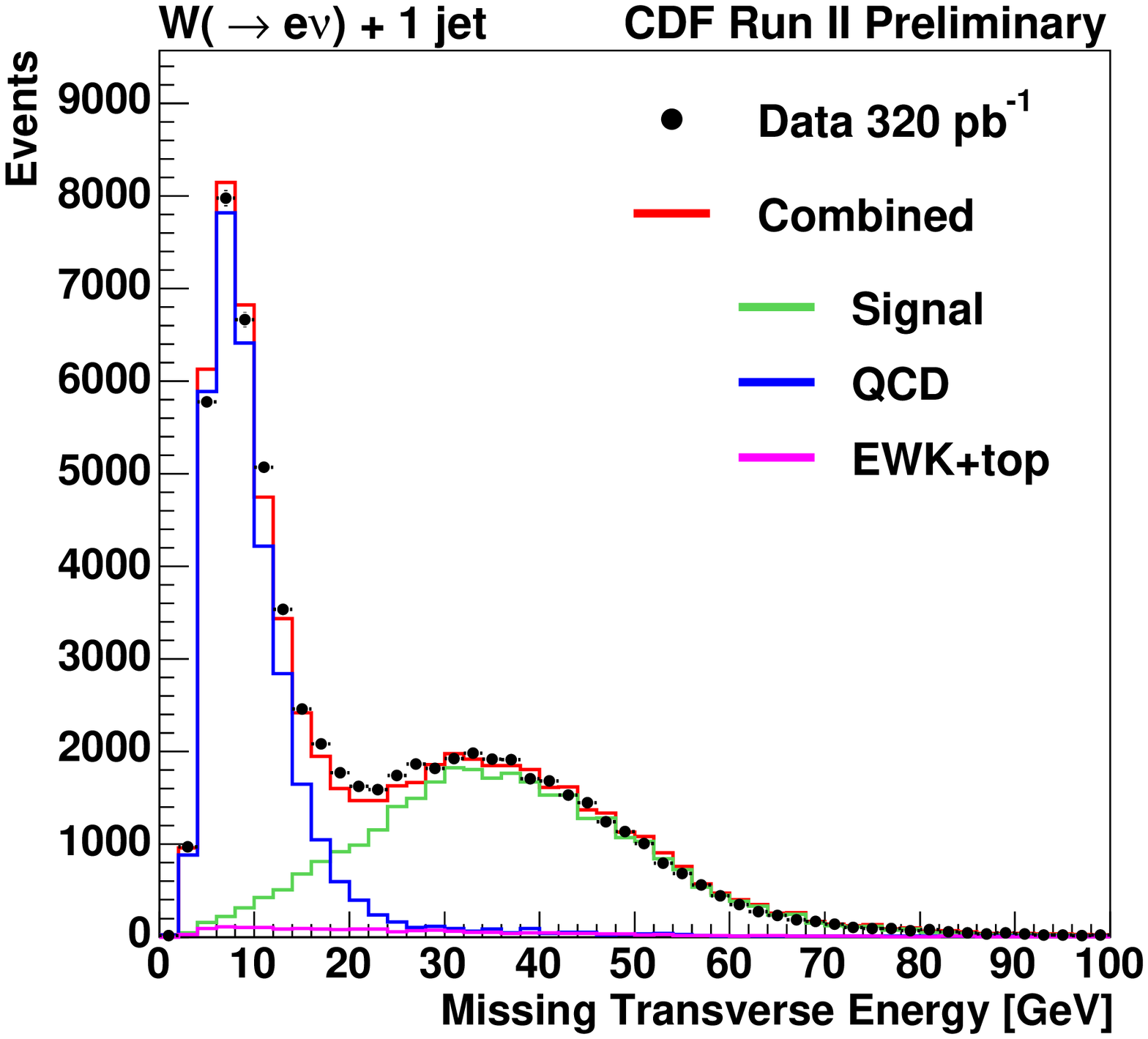}
\includegraphics[width=.181\textheight]{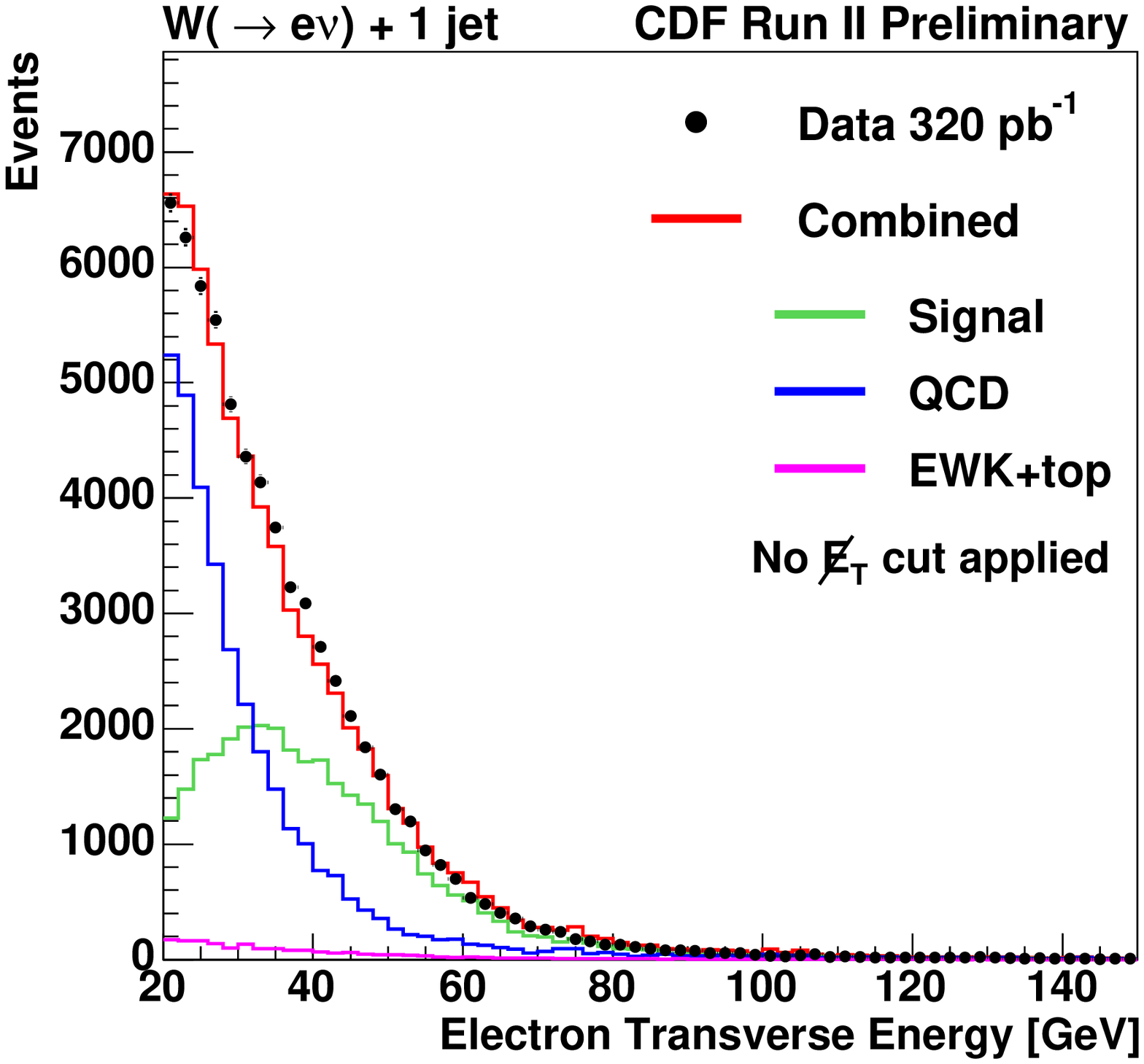}
\includegraphics[width=.181\textheight]{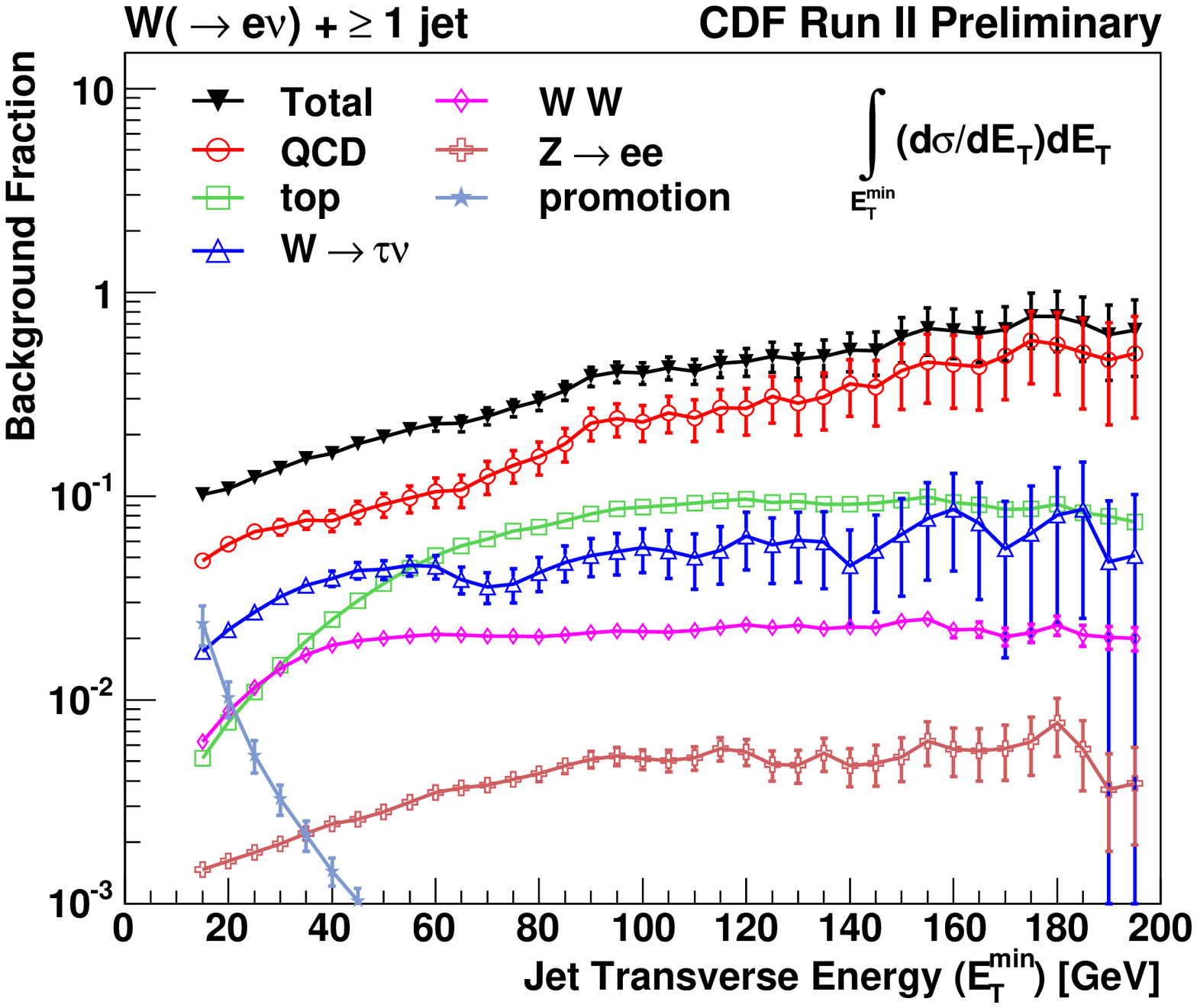}
\includegraphics[width=.181\textheight]{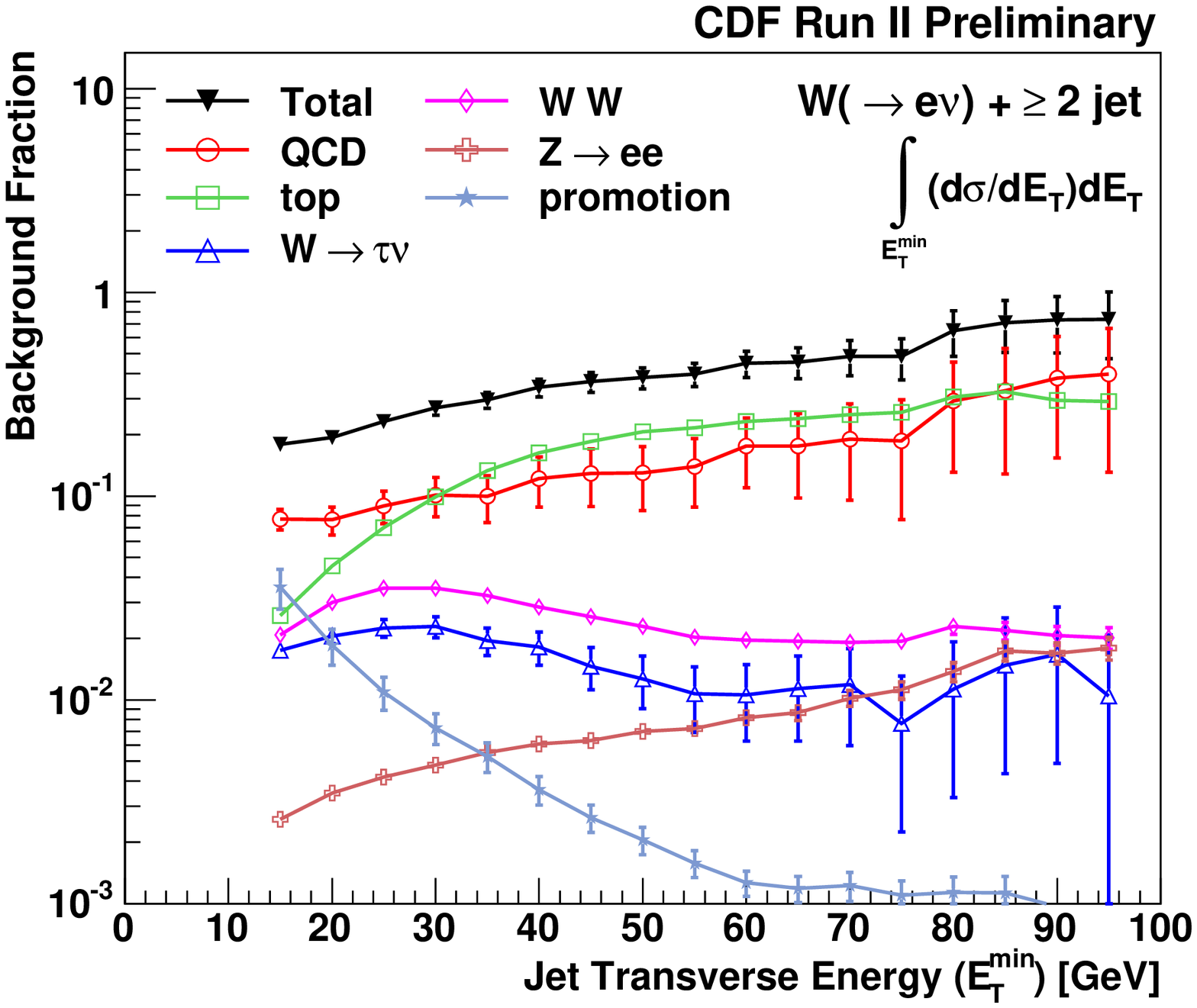}
\caption{
Counter-clockwise: 1) \met ~distribution for event with 1 or more jets. Data are shown in black
along with the templates for the QCD (blue), EWK  (violet) backgrounds and signal (green), the
normalization of each template is determined by the fit to data. 
The red histogram is the sum of the templates resulting from the fit. 2) $E_T^e$ distribution
for events with 1 or more jets. The normalization of each histogram is determined by the fit 
to the \met. 3) and 4) background fraction breakdown as a function of the minimum \etjet, 
respectively for the 2 and 1 jet sample.    
\label{fig:eleet}
}
\end{figure}

Backgrounds to $W+n-{\rm jet}$ production are classified in two categories: QCD and W-like events. 
The latter is represented by events which manifest themselves as real 
electrons and/or \met~ in the final state, namely: $W\to\tau\nu$, $Z\to e^+e^-$,
WW, top pair production. 
The former is mainly coming from jets production in which one or more jets fake an electron 
and have mis-measured energy that results in large \met. 
While the W-like backgrounds are modeled with Monte Carlo
 simulations, the QCD background is described with a data-driven technique.
To extract the background fraction in each $W + \geq n{\rm -jet}$ sample
the  \met ~distribution of candidates is fitted to background
and signal templates (Fig. \ref{fig:eleet} upper left-hand side). 
For this fit the W-like backgrounds and signal are modeled using detector 
simulated Monte Carlo event samples. 
The QCD background is modeled using a ``fake-electron'' event sample, formed
from the same candidate trigger dataset by requiring that at least two of the lepton
identification cuts fail while maintaining all kinematic requirements.
Cross-checks of this method have been performed by looking to other W kinematic 
distributions as the transverse mass of the W $m_T^W$ and the electron $E_T^{e}$ 
(fig. \ref{fig:eleet} upper right-hand side). 
In all these variables a very good agreement between data and background models has been found.

\begin{figure}
\includegraphics[width=.235\textwidth,height=.18\textheight]{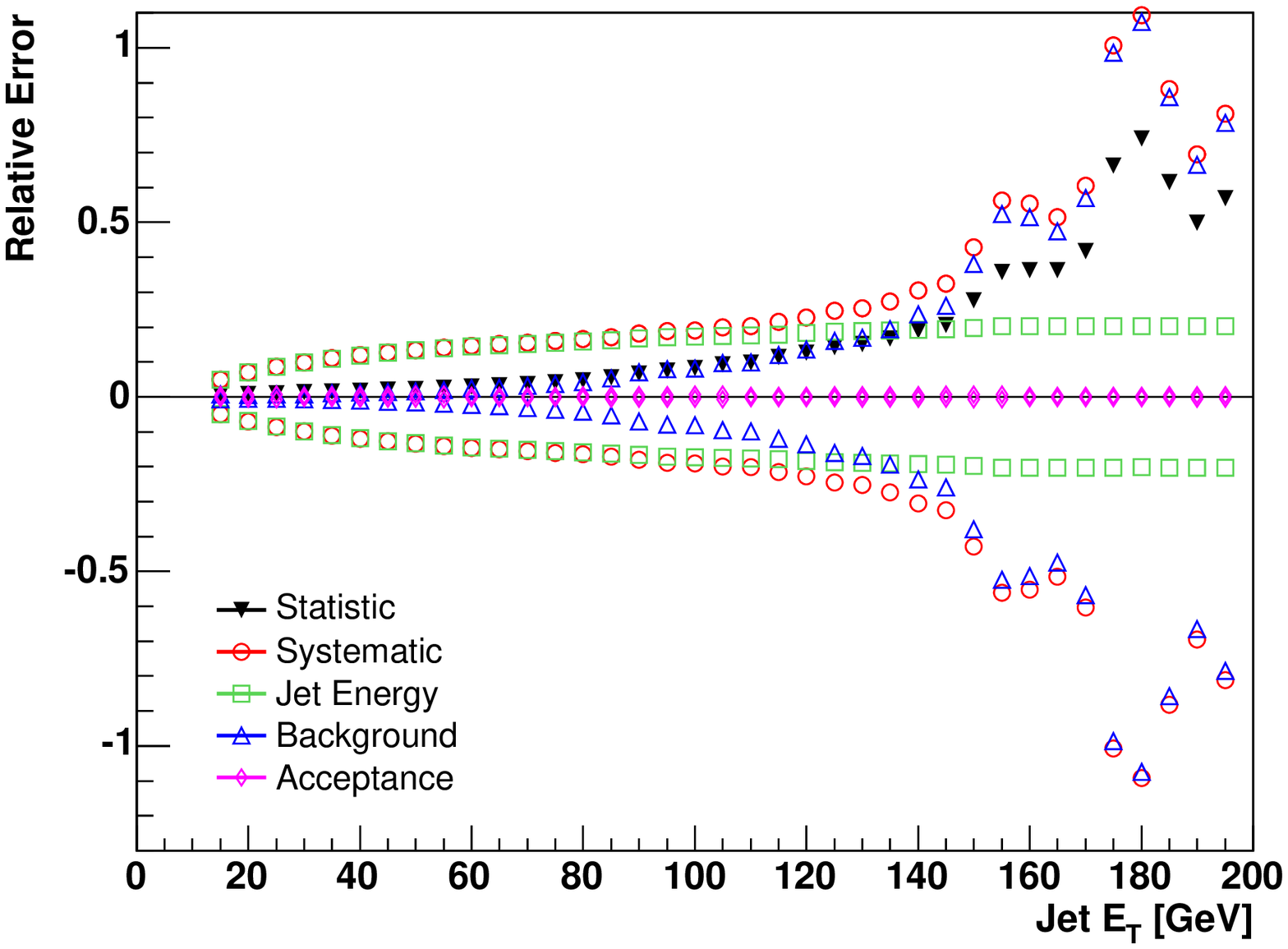}
\includegraphics[width=.235\textwidth,height=.18\textheight]{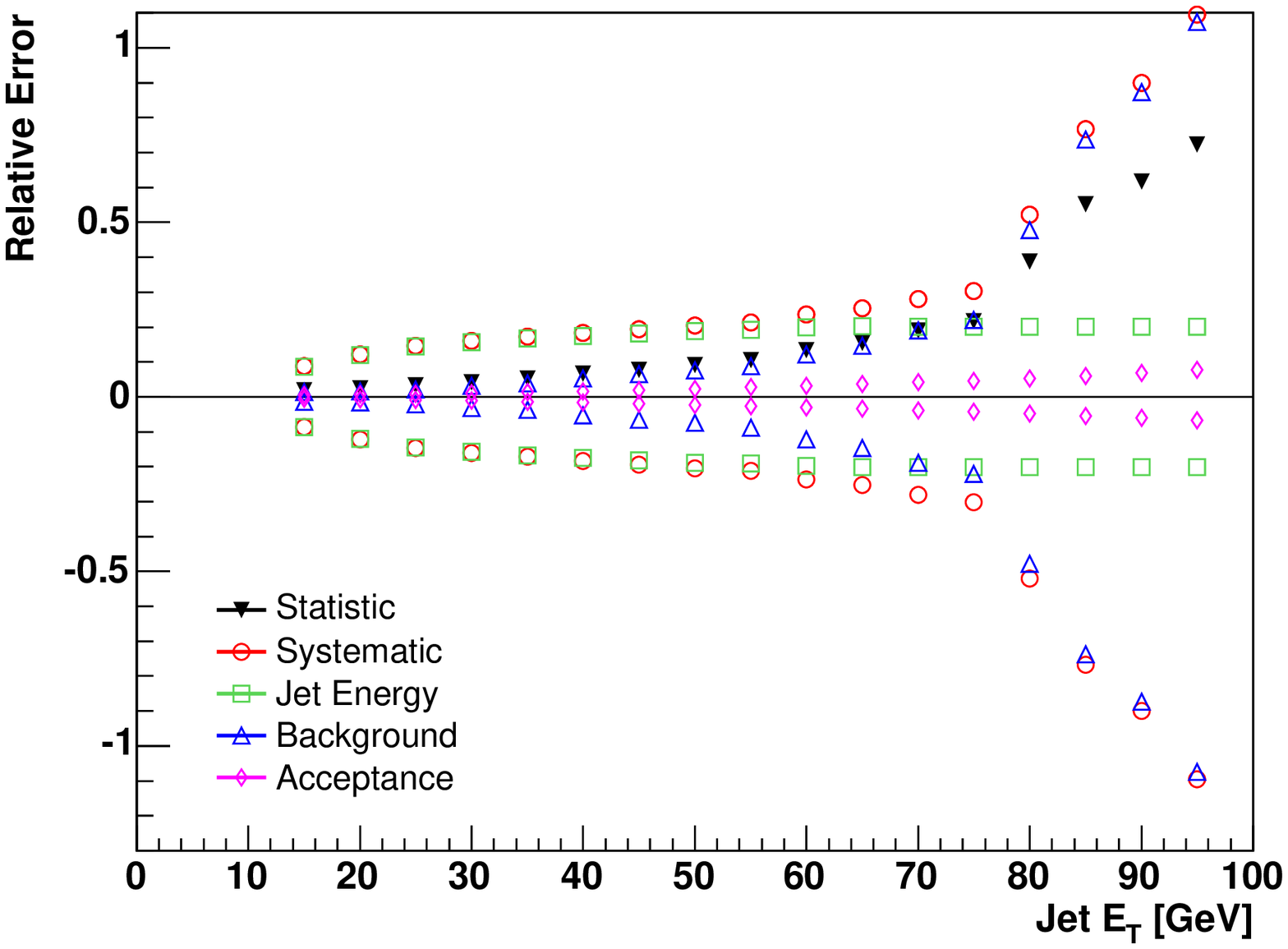}
\caption{Error breakdown for the cumulative 
cross section in the $W+\ge 1 -{\rm jet}$  (left-hand side) and $W+\ge 2 -{\rm jet}$
(right-hand side). The statistic uncertainty is plotted in black, the sum in quadratute
of all the systematic ucertainty in red, the uncertainty associated to the jet energy
scale in green, the uncertainty in the backgraund fraction in blue and the uncertainty
on the acceptance correrection in violet. 
\label{fig:syst}
}
\end{figure}

The total background fraction increases with increasing jet multiplicity and transverse energy.
At low $E_T^{jet}$ it is 10\% (40\%) in the $1-{\rm jet}$ ($4-{\rm jets}$) sample, rising to  90\%
at the highest $E_T^{jet}$. QCD comprises $~70\%$ of the background in the $1-{\rm jet}$ sample. 
At high jet multiplicities and high $E_T^{jet}$ the top contribution becomes increasingly 
important, climbing to 50\% (80\%) of the total background in the $2-{\rm jet}$ ($3,4-{\rm jet}$) sample.
The behavior of the background in the 1 and 2 jet samples is plotted in the lower part of 
fig. \ref{fig:eleet}, where the background fraction is given as a function of the minimum $E_T^{jet}$ used
to define the $W+{\rm jet}$ sample. 
The systematic uncertainty on the background estimate derives mainly from the limited 
statistics of the ``fake-electron'' sample used to model the QCD background, but at high jet multiplicity
the 10\% uncertainty on the measured top pair production cross section is also significant. 
In fig. \ref{fig:syst} is plotted the effect of this uncertainty on the cross section for W + 1 
and 2 jet as a function of the minimum $E_T^{jet}$.

%
A full detector simulation has been used to take into account selection efficiencies, 
coming from geometric acceptance, electron identification and \met~ and $E_T^{e}$ resolution effects.
%
%
The full CDF II detector simulation accurately reproduces electron acceptance and
identification inefficiencies: no evidence of a difference between data and
simulation have been found in the $Z\to e^{+}e^{-}$ sample.
 To minimize the theoretical uncertainty in the extrapolation of the measurement,
 the cross section has been defined for the W phase space 
 accessible by the CDF II detector: $E_T^{e}>20$GeV, $|\eta^{e}|<1.1$,  
 $E\!\!\!/_T>30$GeV and $m_T^{W}>20$GeV/c$^2$. 
This eliminates the dependence on Monte Carlo models to extrapolate 
the visible cross section to the full W phase space.
 Nevertheless Monte Carlo events have been used to correct for inefficiency and boundary
 effects on the kinematic selection that defines the cross section.
 Different Monte Carlo prescriptions
 have been checked and the critical parameters have been
 largely scanned. These effects turned out to be at the 5\% level at low $E_T^{jet}$.
 They have been included into the systematic uncertainty 
 on the efficiency which is $(60\pm3)$\%, largely independent of the jet kitematic. 
As shown in fig. \ref{fig:syst}, the acceptance contribute a  small error to the total uncertainty
on the cross section. 

%
The candidate event yield, background fractions, and acceptance 
factors  are combined to form  the ``raw'' \wgenjet cross section 
in each bin of the \etjet ~spectra. 
The raw cross sections are unfolded for detector effects on the measured jet energies and corrected 
to the hadron level using Monte Carlo events. 
\texttt{Alpgen} \cite{ref:alpgen} interfaced with \texttt{PYTHIA-TUNE A}~\cite{ref:PYTHIA,ref:TUNEA}
provides a 
reasonable description of the jet and underlying event properties, and is used to 
determine the correction factors, defined as the ratio of the hadron level cross section 
to the raw reconstructed cross section, used in the unfolding procedure.
To avoid dependence of such a correction on the assumed Monte Carlo hadron level \etjet distribution,
an iterative procedure is used to reweight the events at the hadron level until the hadron level~\etjet
distribution agrees with the corresponding data-unfolded distribution to within the systematic
uncertainties on the measurement.
The unfolding factors vary between $0.95$ and $1.2$ over the measured range of \etjet. 
The measured jet energies were varied by $\pm\sigma\sim 3\%$ as detailed in \cite{ref:jetnim}, to
account for systematic effects introduced by the uncertainty on the calorimeter 
absolute energy scale.
The total systematic on the cross section introduced by the jet energy measurement is dominated by
the uncertainty on the absolute energy scale and ranges between 5\% and 20\%, increasing with ~\etjet.

\begin{figure*}
\includegraphics[height=.245\textheight]{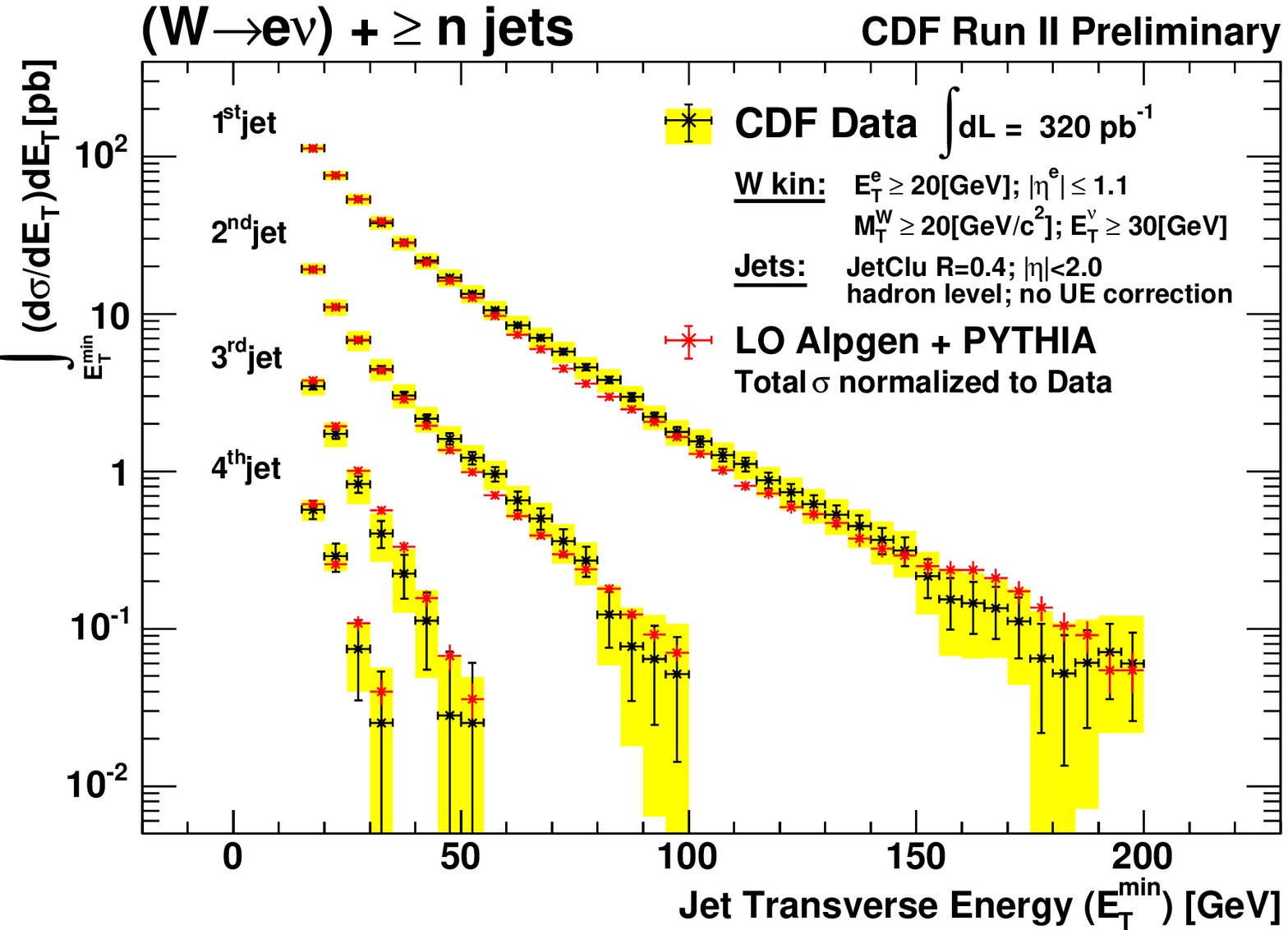}
\includegraphics[height=.245\textheight]{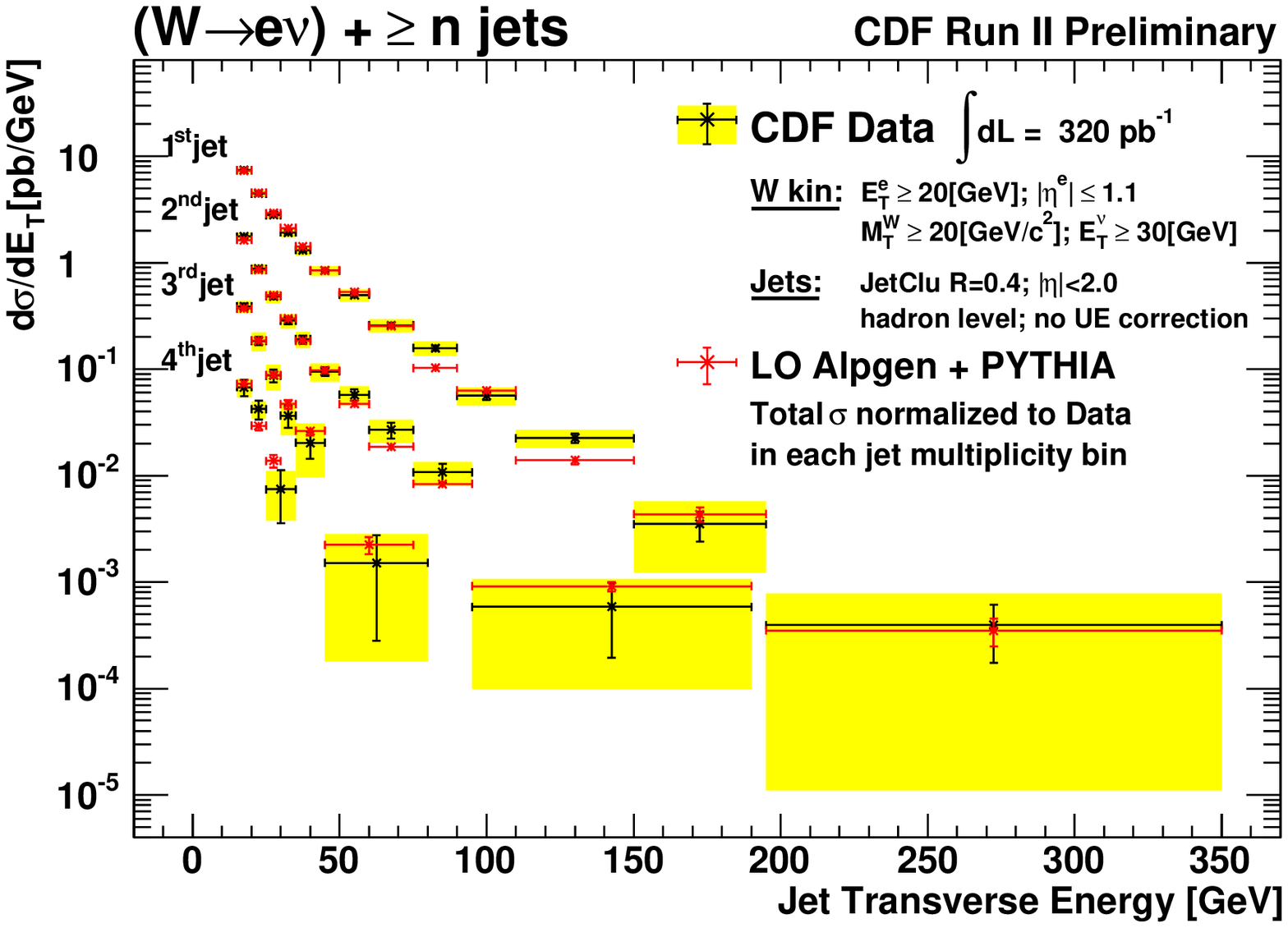}
\includegraphics[height=.245\textheight]{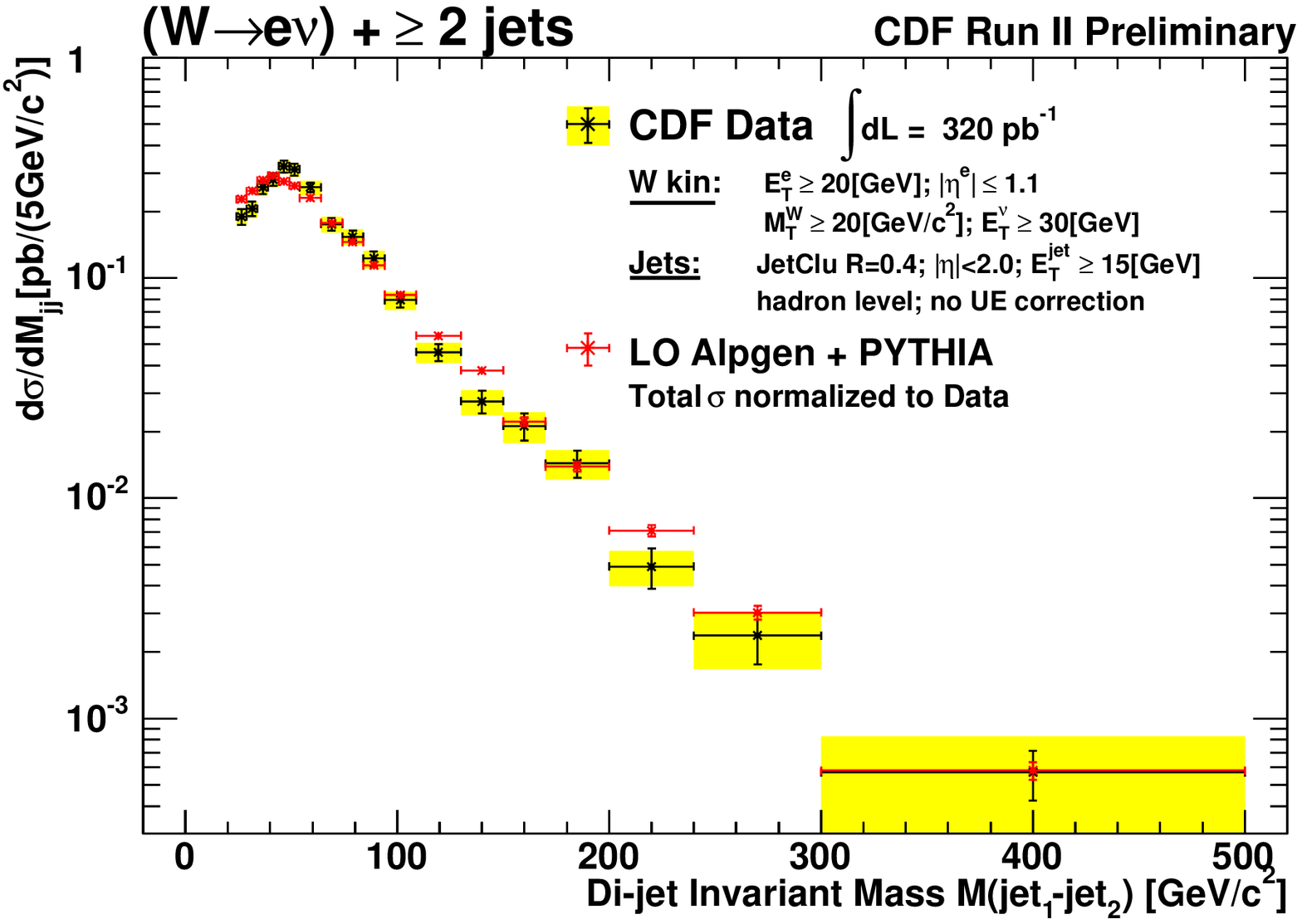}
\includegraphics[height=.248\textheight]{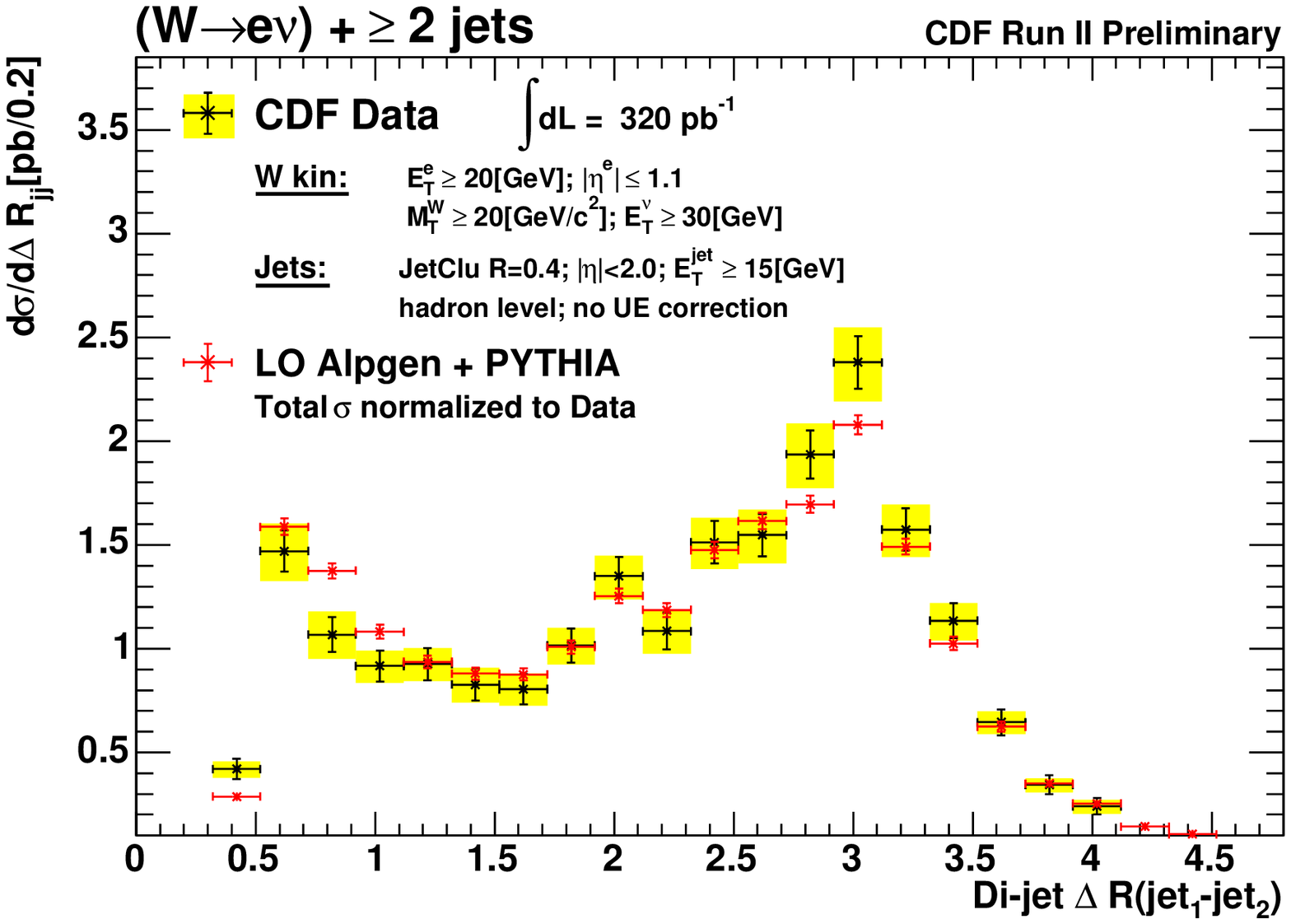}
\caption{Top: Cumulative cross section $\sigma(W\to e\nu+\geq n-{\rm jets}; E_{T}^{jet}(n) > E_{T}^{jet}(min))$
as a function of the minimum $E_T^{jet}(min)$ (Left) and differential cross section 
$d\sigma(W\to e\nu+\geq n-{\rm jets})/d E_{T}^{jet}$ (Right) for the first, second, third and fourth inclusive 
jet sample. Bottom: Differential cross section  $d\sigma(W\to e\nu+\geq 2-{\rm jets})/d M_{j1j2}$ (Left) and
$d\sigma(W\to e\nu+\geq 2-{\rm jets})/d R_{j1j2}$ (Right) respectively as a function of the invariant
mass and angular separation of the leading 2 jets. 
Data are compared to \texttt{Alpgen}+\texttt{PYTHIA} predictions normalized to
the measured cross section in each jet multiplicity sample.
\label{fig:xs}}
\end{figure*}

The measured cross section are shown in fig.~\ref{fig:xs}. Results are presented as both cumulative
$\sigma(W\to e\nu+\geq n-{\rm jets}; E_{T}^{jet}(n) > E_{T}^{jet}(min))$
and differential
$d\sigma(W\to e\nu+\geq n-{\rm jets})/d E_{T}^{jet}$
distribution where $E_{T}^{jet}$ is that of the $n^{th}-$leading jet (upper plots fig.~\ref{fig:xs}).
The measurement spans over three orders of magnitude in cross section and close to 
$200$~${\rm GeV}$ in jet $E_{T}$ for the $\geq 1-{\rm jet}$ sample. For each jet multiplicity, 
the jet spectrum is reasonably well described by
individually normalized \texttt{Apgen}+\texttt{PYTHIA} $W+n-{\rm parton}$ samples.
The shape of the dijet invariant mass and angular correlation (lower plots fig.~\ref{fig:xs}) are also
well modeled by the same theory prediction.
In fig \ref{fig:xs} 
the solid bars represent the statistical uncertainties on the event yield in each bin, while the shaded bands 
are the total systematic uncertainty which is the sum in quadrature of the effects introduced by the uncertainty 
in the background estimation, 
efficiency correction and jet energy measurement (fig. \ref{fig:syst}). The systematic uncertainty is $< 20\%$ at low \etjet 
increasing  to $50\%-100\%$ at high \etjet for all $n-{\rm jet}$ cross sections. 
At low \etjet the systematic 
error is dominated by the uncertainty on the jet energy scale, whereas at high \etjet it is 
dominated by the background uncertainty, 
in particular, by the limited statistic of the QCD background sample. 
We expect to reduce drastically this effect by increasing the statistic of the data sample.
The behavior of the uncertainties as a function of the \etjet ~is similar for both the 
the cumulative and differential cross section. The invariant mass shows a similar increase of the
uncertainty with increasing di-jet masses. The angualr correlation, on the other hand, has an
uncertainty reasonably independent of the $\Delta R$ separation and dominated by the background
subtraction.

In summary, we have measured the \wgenjet cross sections in $320$~${\rm pb^{-1}}$ of $p\bar{p}$ 
collisions at $\sqrt{s}=1.96$ TeV, including events with $4$ or more jets produced in 
association with the $W$ boson. The cross sections, defined in a limited $W$ decay phase space,
have otherwise been fully corrected for all known detector effects. Preliminary comparisons show
reasonable agreement between the measured cross sections and the predictions of matched Monte 
Carlo samples.

\end{document}